\documentclass{aa}  

\usepackage{graphicx}
\usepackage{tablefootnote}
\usepackage{threeparttable}
\usepackage{subfigure}
\usepackage{xcolor}
\usepackage{txfonts}
\usepackage{natbib}
\begin{document} 

   \title{Exploring the non-thermal physics behind the pulsar wind nebula PSR\,J2030$+$4415 through radio observations}

   \author{J.M. Paredes
          \inst{1}
          \and
          P. Benaglia\inst{2}
          \and 
          V. Bosch-Ramon\inst{1}
          \and
          A.Tej\inst{3}
          \and
          A. Saha\inst{3}
          \and
           J. Martí\inst{4}
          \and
          P. Bordas\inst{1}
          }

   \institute{Departament de Física Quàntica i Astrofísica, Institut de Ciències del Cosmos, Universitat de Barcelona, IEEC-UB, Martí i Franquès 1, 08028 Barcelona, Spain\\
              \email{jmparedes@ub.edu}
         \and
             Instituto Argentino de Radioastronomía, CONICET-CICPBA-UNLP, CC5(1897) Villa Elisa, Prov. de Buenos Aires, Argentina\\
             \email{paula@iar-conicet.gov.ar}
        \and
            Indian Institute of Space Science and Technology, Thiruvananthapuram 695 547, Kerala, India\\
            \email{tej@iist.ac.in}
        \and
             Departamento de Física (EPSJ), Universidad de Jaén, Campus Las Lagunillas s/n, A3, E-23071 Jaén, Spain\\
             }

   \date{Received ; accepted }

 
  \abstract
   {PSR\,J2030$+$4415 is a gamma-ray pulsar with an X-ray pulsar wind nebula elongated along the north--south direction. The system shows a prominent X-ray filament oriented at an angle of 130° to the nebula axis.  }
   {To improve our understanding of the non-thermal processes occurring in the pulsar wind nebula, we attempted to determine the possible existence of a radio counterpart, study its morphology, and obtain restrictive upper limits of the pulsar and filament emission at radio wavelengths.}
   {We   performed observations of the pulsar PSR\,J2030$+$4415 and its surroundings with the upgraded Giant Metrewave Radio Telescope (uGMRT) at two frequency bands, and put the results in context with findings at other wavelengths.}
   {We   obtained radio images at 736 and 1274 MHz that reveal a structure trailing the pulsar, with a morphology overlapping the X-ray nebula. This radio structure is the radio counterpart of the X-ray pulsar wind nebula. The derived spectral index along this structure shows spatial variation. There are no hints of the pulsar and the filament at any of the explored radio frequencies, but we  obtained restrictive upper limits. A physical scenario that combines the radio and the X-ray observations, and consistent with IR data, of the nebula and the filament is presented. We propose that particle acceleration occurs in the nebula tail due to the presence of a re-collimation shock, and the highest energy particles gradually escape from it through energy-dependent diffusion. We also find a lower limit in the energy of the particles escaping along the X-ray filament of $\sim$~GeV.}
   {}

               
 \titlerunning{The pulsar wind nebula PSR\,J2030$+$4415}
 \authorrunning{Paredes et al.} 
 
   \maketitle
   
%

\section{Introduction}\label{intro}
PSR\,J2030$+$4415 is a radio quiet pulsar discovered in a blind search of \textit{Fermi} Large Area Telescope (LAT) data (\citealt{Pletsch2012}). The recently published Third \textit{Fermi} Large Area Telescope Catalog of Gamma-Ray Pulsars (\citealt{Smith2023}) lists a total of 294 gamma-ray pulsars that includes PSR\,J2030$+$4415, for which they give a pulsar period of $P = 227.1$~ms, its first derivative $\dot{P}= 5.05\times 10^{-15}$, and a spin-down luminosity of $\dot{E}=1.7\times 10^{34}$~erg~s$^{-1}$. The derived characteristic age is  $\tau = P/2\dot{P}= 7.13\times10^5$~yr. The pulsar is moving with a transverse velocity $v_{\perp}=404\pm 86\,d_{\rm kpc}$~km~s$^{-1}$ and with a position angle PA$=7.2^{\circ}\pm7.6^{\circ}$ (\citealt{Vries2020}). The distance to the pulsar is 
$\sim$ 0.5~kpc (\citealt{Vries2022}). 

H$_{\alpha}$ images of PSR\,J2030$+$4415 show a very bright compact bow shock coincident with the pulsar and a region of extended diffuse emission formed by two bubbles trailing the pulsar \citep{Brownsberger2014}. 
A third fainter bubble, surrounding the pulsar, has been discovered by \cite{Vries2020}.

Observations of PSR\,J2030$+$4415 in the X-ray band have revealed a complex system morphology, including an unresolved X-ray source coincident with the pulsar, a trailing pulsar wind nebula (PWN), and a prominent X-ray jet-like structure (\citealt{Marelli2015}; \citealt{Vries2020,Vries2022}). 

At radio frequencies, PWNe are characterized by typical spectral indices $-0.3\lesssim\alpha\lesssim0$, where $S_{\nu} \sim \nu^{\alpha}$ and $S_{\nu}$ is the flux density at frequency $\nu$. In X-rays, the spectra of PWNe are steeper, with    photon indices in the range $1 \lesssim \Gamma \lesssim 2$ and with a median of $\Gamma = 1.7$, and $\Gamma \equiv 1-\alpha$ (\citealt{Kargaltsev2008}). PWNe interacting with the surrounding interstellar medium (ISM) can give rise to distinct extended morphological features. Torus-like structures and bipolar jets are observed in pulsars displaying slow proper motions, as well as bow-shaped shocks and extended cometary tails in pulsars moving at high velocities  (see e.g. \citealt{Gaensler2006}). 
In the X-ray band, these structures have been resolved in detail in several cases (e.g. \citealt{Kargaltsev2008}). The shock formed at the termination of the pulsar wind accelerates particles (electrons and positrons) that are advected by the shocked flow in the opposite direction to the pulsar motion. These relativistic particles are responsible for the X-ray PWN powered by PSR\, J2030$+$4415 and observed with {\it Chandra}, and lower energy electrons and positrons from the same non-thermal particle population are responsible for the radio counterpart. 

PSR\,J2030$+$4415 
belongs to a small subset of supersonically moving PWNs (sPWNs) that exhibit unusually long X-ray jet-like structures. To date, only five secure filaments and three candidates have been identified (\citealt{Dinsmore2024}). Some of the most prominent examples of such structures are found in the Guitar Nebula (\citealt{Hui2012}) and in the Lighthouse Nebula (\citealt{Pavan2014}), but the number of  detection of large-scale filaments from sPWNe is gradually increasing,   for example the recent findings in PSR\,J1135$-$6055 (\citealt{Bordas2020}) and J1809--1917 (\citealt{Klingler2020}). In none of these sources has the filament been detected in radio. Very recently, however, \cite{Khabibullin2024} reported the discovery of a faint one-sided radio-emitting filament located ahead of the pulsar PSR\,J0538$+$2817, in the direction of the pulsar’s proper motion, with no detected X-ray or optical emission. To explain the first discovered X-ray filament found in the Guitar Nebula, \cite{Bandiera2008} proposed a scenario where the most energetic particles that are accelerated at the pulsar wind termination shock may escape from this region and diffuse into the predominantly ordered magnetic field of the surrounding medium, giving rise to synchrotron X-ray emission. 
 
To determine the possible existence of the radio PWN in PSR\,J2030$+$4415 and study its morphology, we performed observations with the upgraded GMRT (uGMRT) and used archival X-ray data from {\it Chandra} and infrared data from WISE to place the radio observations in context. We report here the discovery of the radio counterpart of the X-ray PWN associated with this pulsar and a restrictive upper limit for the undetected pulsar and X-ray filament. The WISE infrared images show a lack of emission that matches the PWN morphology. A physical scenario able to explain the available multi-wavelength data is discussed.

   \begin{figure}
\centering
   \includegraphics[width=0.497\textwidth]{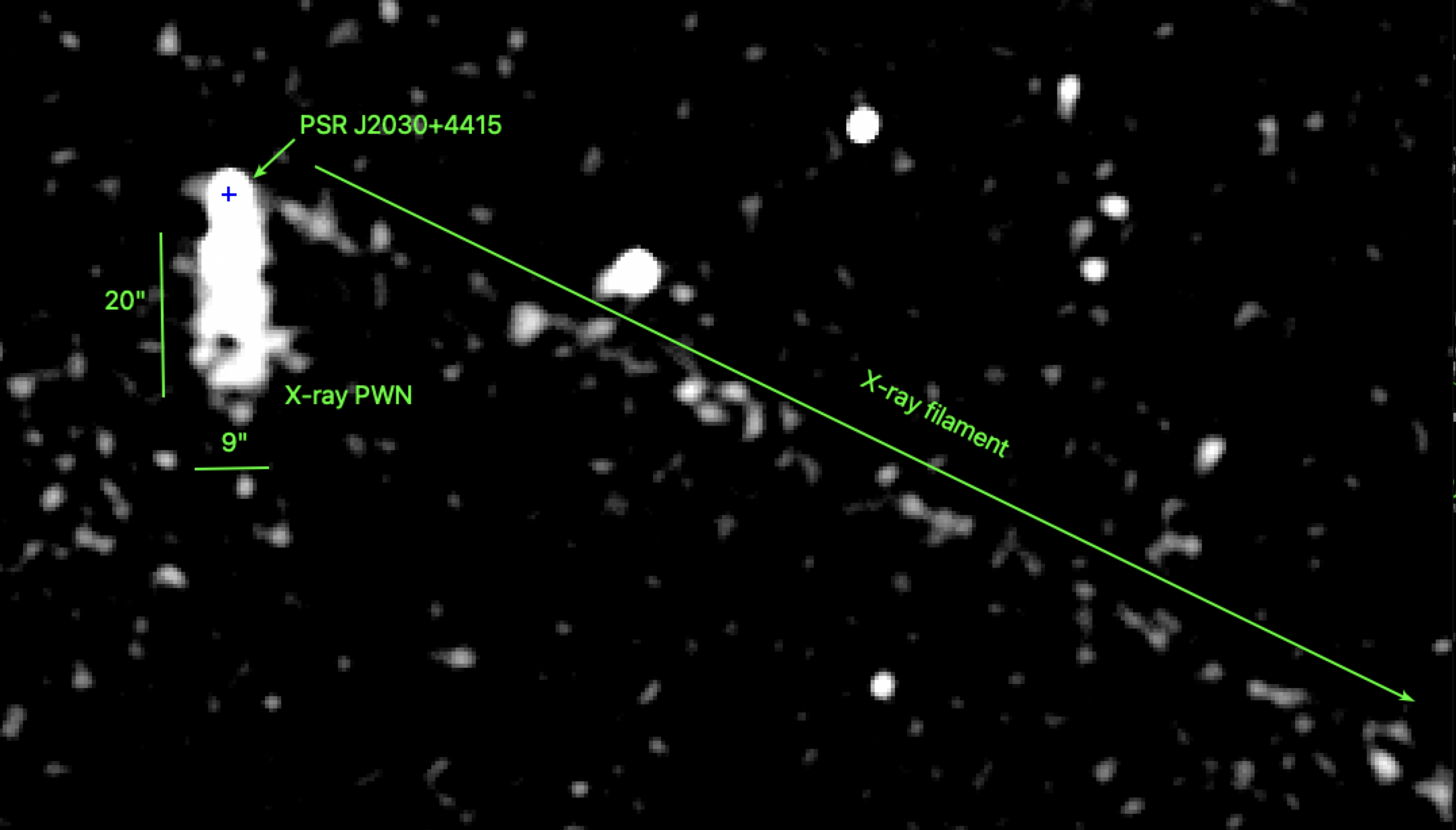} 
      \caption{Archive {\it Chandra} image of the PSR\,J2030$+$4415 field in the 0.5--7 keV energy range. Image adapted from the original   produced by \cite{Vries2022} after merging several observations (Obs. Id. 14827, 20298, 22171-22173, 23536, 24954, 24236) obtained from April 2014 to November 2021. The blue cross marks the position of PSR\,J2030$+$4415. In the image we show only a fraction (160\arcsec) of the large-scale filament ($15\arcmin$) \citep{Vries2022}. The length and the width of the trailing PWN are 20\arcsec and 9\arcsec, respectively.  North is up and east to the left.
    }
         \label{FigFil}
   \end{figure}
%

\section{X-ray data}

{\it Chandra} and {\it XMM-Newton} observations of the $\gamma$-ray pulsar PSR\,J2030$+$4415 identified its X-ray counterpart at R.A.(J2000) = 20h~30m~51.4s ($\pm0.15\arcsec$) and DEC(J2000)= $44^{\circ}15\arcmin38.8\arcsec$ ($\pm0.16\arcsec$), with a significance of 15$\sigma$, and provided evidence for a $\sim 10\arcsec$-long PWN elongated along the north-south direction (\citealt{Marelli2015}). 
The 0.3--10 keV spectrum of the tail is well described by a power law, with a photon index $\Gamma_{\rm X, PWN} = 1.2^{+0.5}_{-0.4}$, with the unabsorbed X-ray flux of $F_{\rm X, PWN} = (4.0 \pm 1.6) \times 10^{-14}$~erg~cm$^{-2}$ s$^{-1}$ (\citealt{Marelli2015}). At a distance of 0.5\,kpc this gives an X-ray luminosity of $L_{\rm X, PWN}=1.2\times 10^{30}$~erg~s$^{-1}$.
Later observations by {\it Chandra} revealed an unresolved X-ray source coincident with the pulsar, and confirmed the PWN tail, which extended to $\sim 20\arcsec$, as well as the presence of a one-sided filament extending into the surrounding medium for at least 5\arcmin\ \citep{Vries2020}. 
The spectral fit of the PWN tail gives $\Gamma_{\rm X, PWN} = 1.48\pm0.10$, a hydrogen column density $N_{\rm H}=0.6\times10^{21}$cm$^{-2}$, and a total X-ray flux of $\sim 3 \times 10^{30}$ erg s$^{-1}$ using a distance of 0.5 kpc. The fits along the PWN tail 
showed a softening with distance from the pulsar position, with $\Gamma_{\rm X, PWN}$ varying from 1.31 to 2.17. This softening may reflect significant synchrotron cooling of the particles during the $\sim 300$ years that the pulsar has taken to traverse this structure \citep{Vries2020}.
As discussed later, this timescale would imply a bulk velocity of the emitting flow similar to that of the pulsar, and much lower than the shocked pulsar wind filling the PWN tail.
Deeper {\it Chandra} observations showed an extension of the filament of more than 15\arcmin (2.2~pc), with a typical filament width of 4\arcsec, and oriented at an angle $130^{\circ}$ to the PWN axis \citep{Vries2022}. These authors made a spectral fit of the filament with an absorbed power-law model ($N_{\rm H}=0.6\times10^{21}$cm$^{-2}$) and obtained $\Gamma_{\rm X, Fil} = 1.50\pm0.12$ and a 0.5–7 keV unabsorbed flux per arcmin filament length of $8.0^{+0.7}_{-0.6} \times 10^{-15}$~erg~cm$^{-2}$ s$^{-1}$ arcmin$^{-1}$.

We   used a {\it Chandra} archive image, originally produced by \cite{Vries2022} after merging several observations (Obs. Id. 14827, 20298, 22171-22173, 23536, 24954, 24236) obtained from April 2014 to November 2021, to reproduce in Fig.~\ref{FigFil} the X-ray field around PSR\,J2030$+$4415.

\section{Previous radio observations towards PSR\,J2030$+$4415}
We searched for any hint of emission at 1.4 GHz 
in the NRAO VLA Sky Survey (NVSS), the Very Large Array Sky Survey (VLASS), and the Rapid ASKAP Continuum Survey (RACS) \citep[][respectively]{Condon1998,Lacy2020,Duchesne2023}.
A discrete source, centred at the field of view (FoV) centres of the present observations, was detected in the RACS-mid image (1367~MHz, synthesized beam of $40.4'' \times 11.2''$) with a flux density of $2.9\pm0.5$~mJy. Due to its extension and that of the synthesized beam, it encompasses the location of PSR\,J2030$+$4415 at one of its ends.

At lower frequencies of 325 MHz and 610 MHz,
\citet{Benaglia2020,Benaglia2021} performed a deep survey of the Cygnus region with the GMRT, including PSR\,J2030$+$4415 and its nearby surroundings.  We   used these data to study in detail the radio emission in the field of PSR\,J2030$+$4415. This study   reveals a source at 610 MHz of $2.2\pm 0.13$ mJy, with a size of about $15\arcsec$ and located $8\arcsec$ south of the pulsar position. This radio source partially  overlaps the southern extension of the X-ray PWN and is likely the radio counterpart. This source was not detected in the 325 MHz image, giving an upper limit of 1.8 mJy beam$^{-1}$ at a $3\sigma$ confidence level. 



\section{GMRT Observations}

We performed observations of the pulsar PSR\,J2030$+$4415 and its surroundings with the uGMRT on June 29, 2023, and on June 30 (Project code: 44\_014). Specifications of the GMRT systems are given in \citet{Swarup1991} and \citet{Gupta2017}. 

We observed in two frequency bands, 
Band 4 (550-950 MHz, 5~h) and Band 5 (1050-1450 MHz, 4~h), using the GMRT Wideband Backend (GWB) correlator with a bandwidth of 400~MHz (4096 channels).
The number of working antennas was 29 in the first run and 30 in the second run. The integration time was 8.053 seconds. The  FoV of the GMRT for the different bands is 38$\pm$3$'$ and 23$\pm$1$'$ at  Band 4 and Band 5, respectively.\footnote{The GMRT: System Parameters and Current Status, \url{http://www.gmrt.ncra.tifr.res.in/doc/GMRT_specs.pdf}} 
%
The 
primary calibrators were 3C286 for Band 4 and 3C48 for Band 5. 
The source 2052$+$365 was used as phase calibrator with 5~min scans, bracketing target scans of 35 or 30 min. 

The data were processed using the CAPTURE\footnote{\url{https://github.com/ruta-k/CAPTURE-CASA6}} continuum imaging pipeline \citep{Kale2021}, which is based on tasks from the Common Software Astronomy Applications \citep[CASA,][]{McMullin2007}. The data reduction process included flagging, calibration, imaging, and self-calibration. The flux density calibration was performed using the \citet{Perley2017} scale. The Multiterm Multi-Frequency Synthesis (MTMFS) algorithm was set in the {\sl tclean} task to minimize deconvolution errors in broad-band imaging \citep{Rau2011}.  Seven rounds of self-calibration were performed, the last one for phase and amplitude. The images were primary beam corrected.

The synthesized beams of the final images at the highest angular resolution were $4.95'' \times 3.22''$ for Band 4 and $2.01'' \times 2.24''$ for Band 5. 
The final bandwidths resulted in 332~MHz at both observation bands, and the centre frequencies were 736~MHz and 1274~MHz.
The attained rms values were up to a 0.040~mJy~beam$^{-1}$ (Band 4) and 0.015~mJy~beam$^{-1}$ (Band 5) on average. 
Although GMRT is capable of measuring polarization, in our case the observations were carried out in the total intensity mode and not in the full polar configuration. Therefore, in the current data, the polarization information is not available.




\section{Results}
\subsection{Extended radio emission: Radio counterpart of the X-ray PWN PSR\,J2030$+$4415} 
\begin{figure*}[ht]
  \centering
\includegraphics[width=0.497\textwidth]{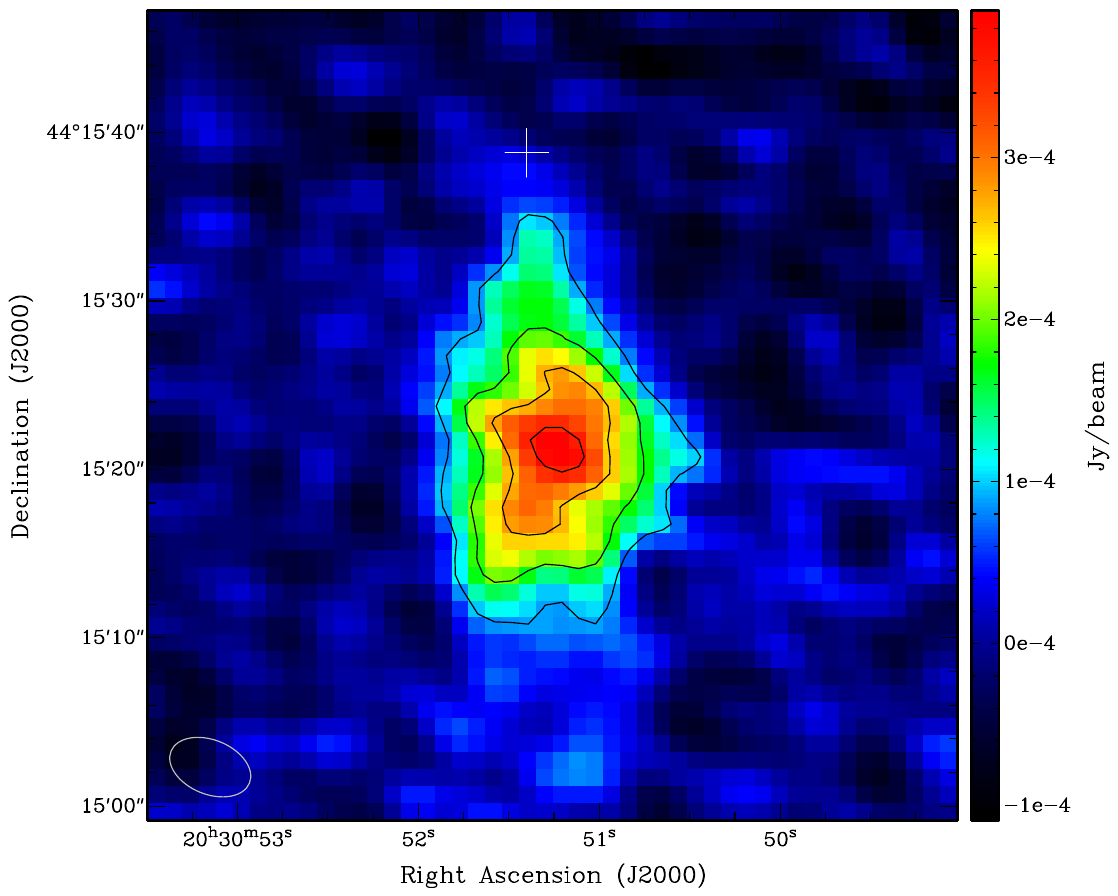}
\includegraphics[width=0.497\textwidth]{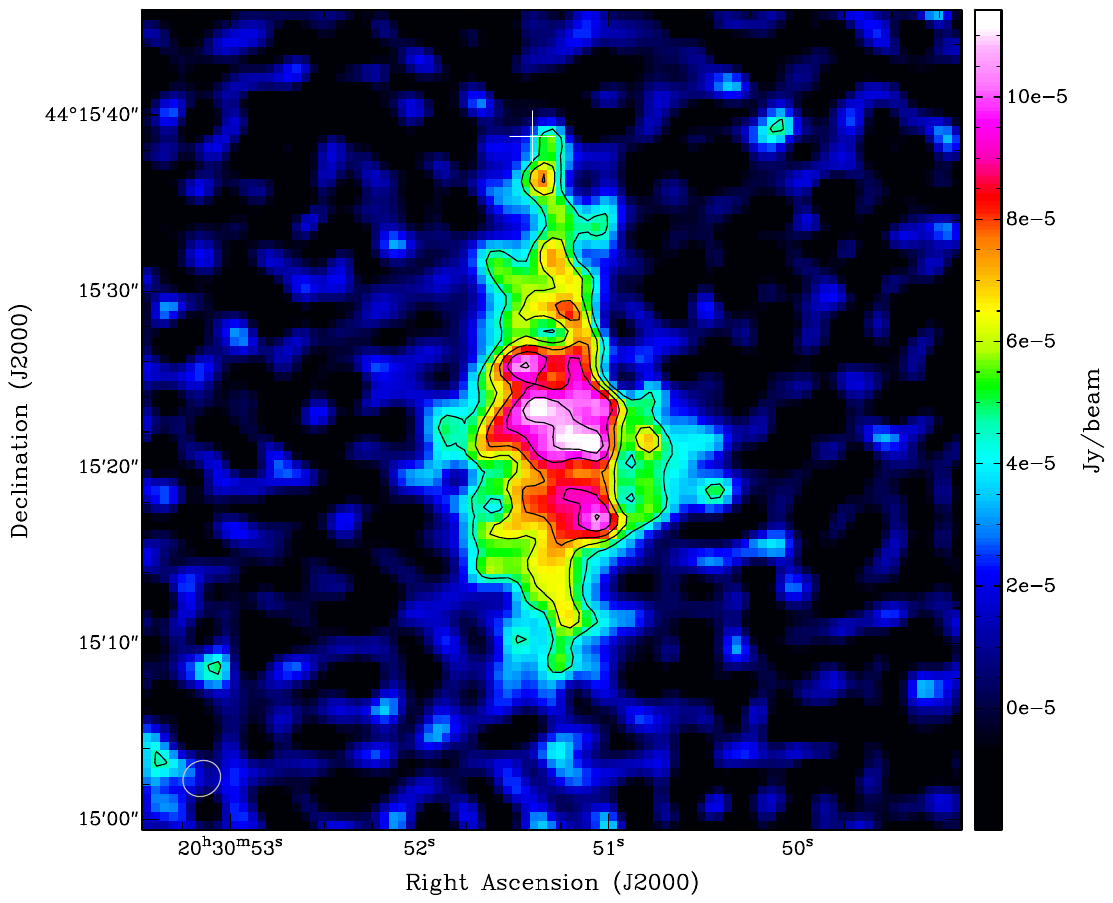}  
  \caption{uGMRT radio images of the PSR\,J2030$+$4415 field. The white cross in both images marks the position of the pulsar PSR\,J2030$+$4415. Left: 736~MHz image, with the synthesized beam shown in the bottom left. Overlaid are the contour levels at 3, 6, 9, and 12 times the rms noise (0.03~mJy~beam$^{-1}$). Right: 1274~MHz image, with the synthesized beam shown in the bottom left. Overlaid are the contour levels at 3, 4, 5, 6, and 7 times the rms noise (0.015~mJy~beam$^{-1}$).}
  \label{FigRadio}
  \end{figure*}

The uGMRT radio continuum images are shown in Fig.~\ref{FigRadio}.
The 736~MHz image shows an extended structure of about $12'' \times 23''$, stretching from north to south. This structure would have a physical dimension of 0.030~pc $\times$ 0.057~pc, 
assuming a distance of 0.5~kpc. The brightness of the radio tail increases with the distance from the pulsar, peaking around  $18''$ (0.045~pc) 
south of the pulsar (see left panel of Fig.~\ref{FigRadio}). The entire radio structure has a total flux density of $2.99\pm0.37$~mJy within the 3$\sigma$ contour. 

The 1274~MHz image shows a similar morphology, but a bit narrower and slightly more elongated, $9'' \times 26''$ (i.e. 0.020~pc $\times$ 0.063~pc). The peak emission is at the same position as that of 736~MHz (see right panel of Fig.~\ref{FigRadio}). The flux density within the 3$\sigma$ contour is $3.18\pm0.41$~mJy.

\begin{figure}
   \centering
   \includegraphics[width=0.42\textwidth, angle=0]{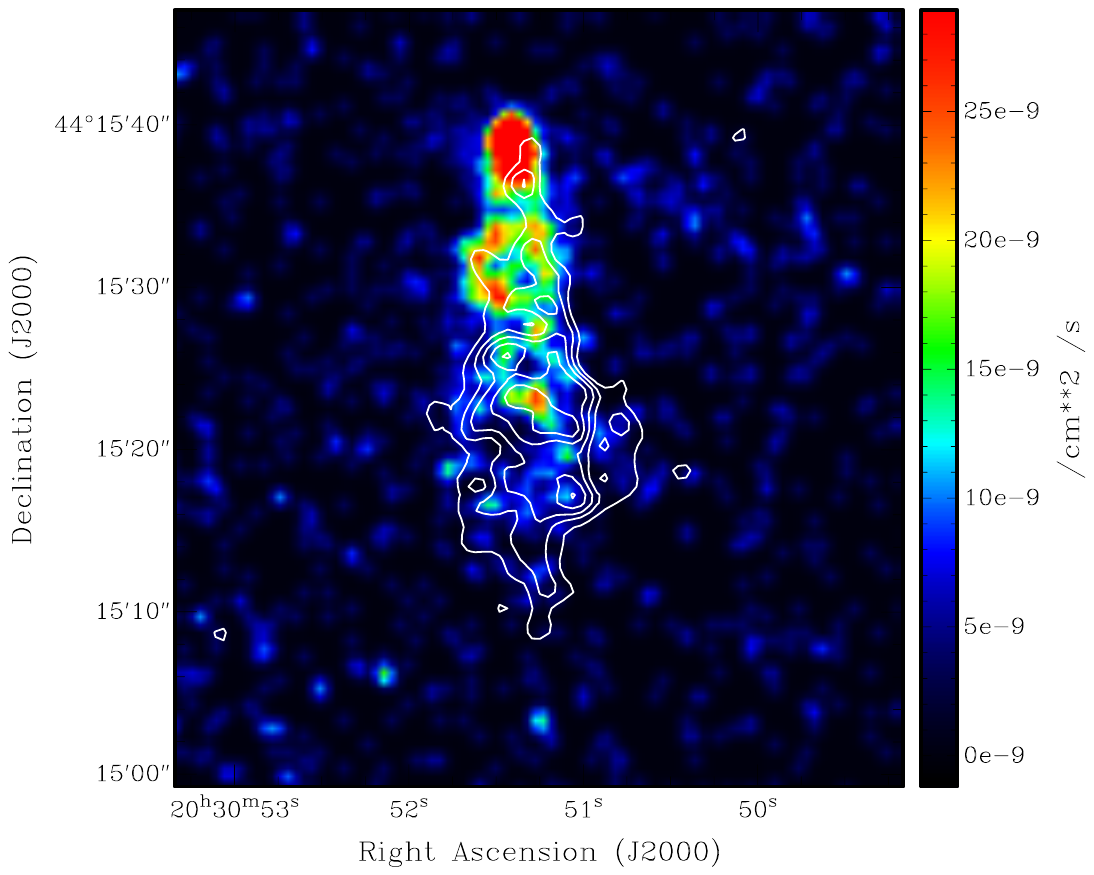}
\caption{Archive {\it Chandra} image of the PWN PSR\,J2030$+$4415 convolved to a beam of $1\arcsec \times 1\arcsec$. The overlaid white contours represent the 1274~MHz radio emission from GMRT. The contour levels are 3, 4, 5, 6, and 7 times the rms noise (0.015~mJy~beam$^{-1}$).  The white cross marks the position of the pulsar PSR\,J2030$+$4415.}
         \label{FigRX}
   \end{figure}

The total intensity radio images at both frequencies reveal a structure trailing the pulsar. The morphology of this radio structure resembles that of PWNe. The positional coincidence between the pulsar position and one extreme of this elongated radio structure suggests a physical connection between them. This is supported by the proper motion of the pulsar, which is northward and opposite to the position of the radio structure. 
In addition, the radio structure overlaps with the X-ray PWN at both frequencies.
 Figure~\ref{FigRX} illustrates the superposition of X-ray emission and radio emission at the best angular resolution (i.e. Band 5 image), where the CXO image has been smoothed at $1''$. Two main features can be appreciated. First, the strongest radio contour corresponds to enhanced X-ray emission, reinforcing the hypothesis of a physical association between the two objects, and second, the radio tail behind the pulsar is more extended than its X-ray counterpart. This larger extension in radio may be due to the relativistic electron synchrotron ageing effect, where low-energy electrons lose energy at a slower rate, resulting in greater longevity that allows them to travel farther from the pulsar. An additional process discussed below that also leads to this effect is the diffusive escape of the most energetic particles, which depletes the nebula tail of X-ray emitting electrons and positrons.
We conclude that the radio structure we found is the radio counterpart of the X-ray PWN.

\subsection{Radio continuum spectrum of PWN PSR\,J2030$+$4415}

We proceeded to derive the spectral index distribution map of the extended source found between the two observation bands, centred at 736 and 1274~MHz. For this, we re-imaged the two data sets, by keeping the same $uv$-range for both, to a common synthesized beam of $6\arcsec \times 6\arcsec$. As before, seven rounds of self-calibration were required, and the images were primary beam corrected. For the pixel-wise estimation of the spectral index, we corrected for any astrometric differences by carefully aligning the images. 

\begin{figure*}
  \centering
\includegraphics[width=0.497\textwidth]{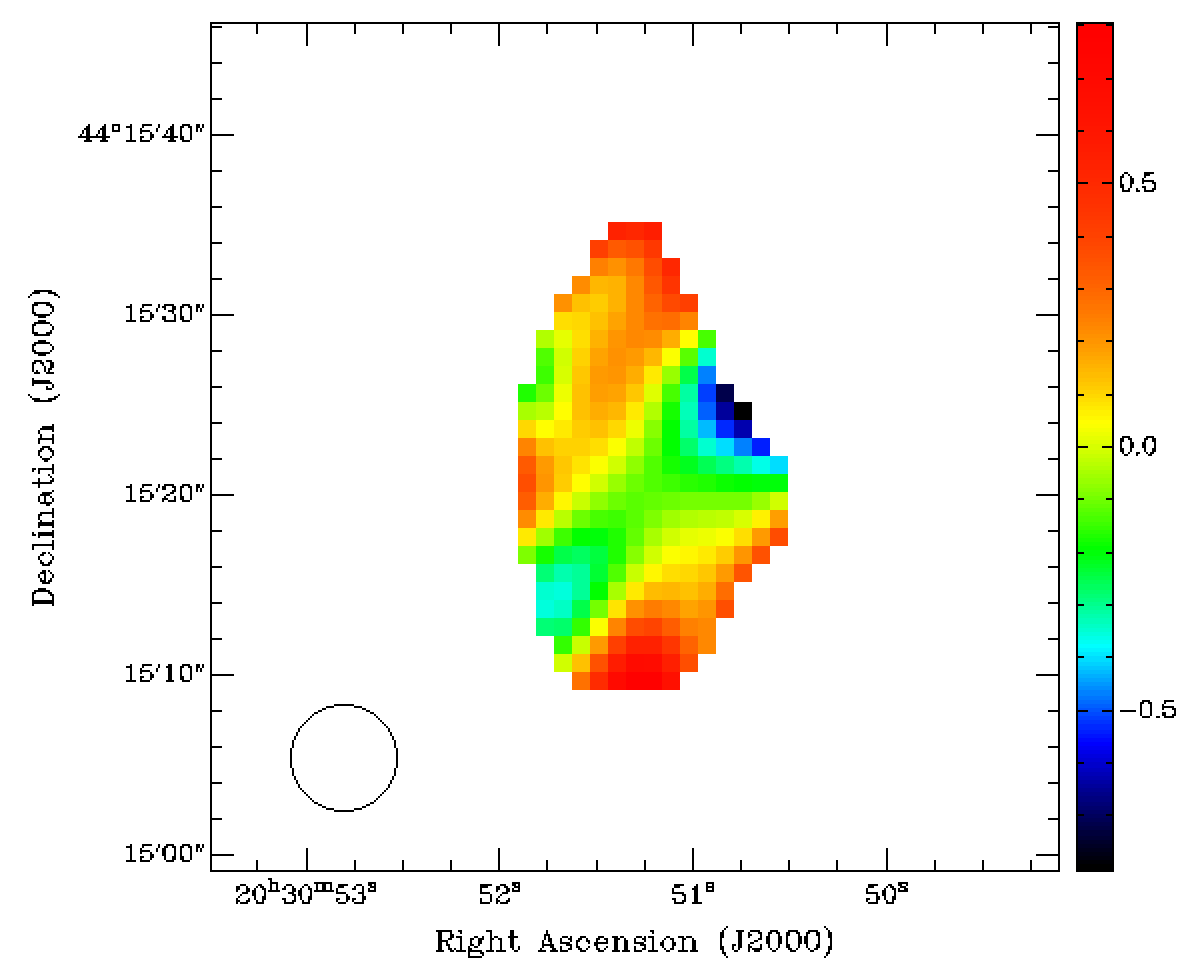}
\includegraphics[width=0.497\textwidth]{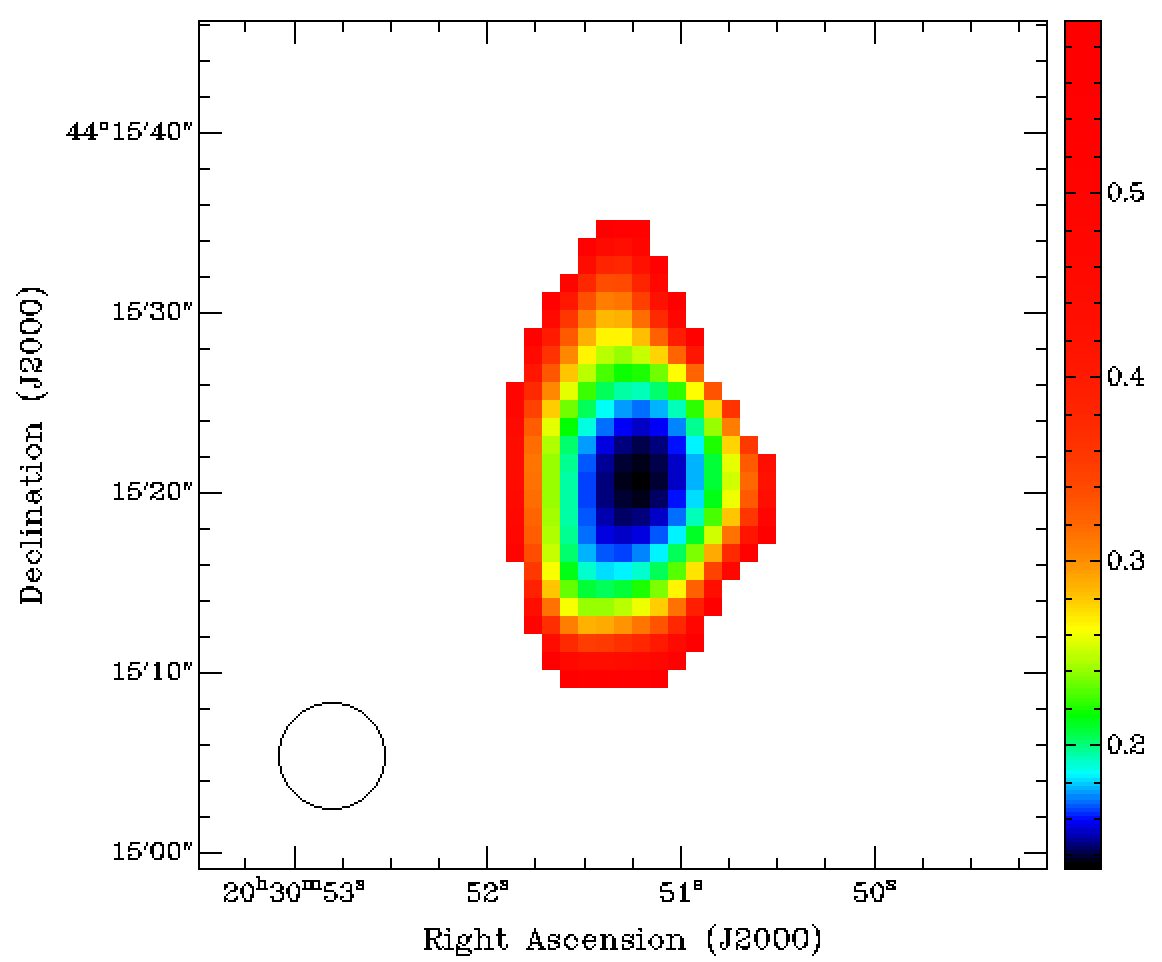}  
  \caption{Distribution of the spectral index and its error along the PWN. Left: Spectral index image of the PWN PSR\,J2030$+$4415 using the Band 4 and Band 5 data. The common synthesized beam of $6\arcsec \times 6\arcsec$ is shown in the bottom left. Right: Spectral index error distribution of the PWN PSR\,J2030$+$4415 using the Band 4 and Band 5 data. The common synthesized beam of $6\arcsec \times 6\arcsec$ is shown in the bottom left.}
  \label{FigIndex}
  \end{figure*}

The spectral index and spectral index error maps were created using the {\sl Miriad} package \citep{Sault1995}. We applied a mask to include only pixels with a signal greater than or equal to 5~rms. In Fig.~\ref{FigIndex} we show the distribution of $\alpha$ and its error along the PWN. There is spatial variation, with $\alpha$ ranging from $-0.60$ to 0.80. 
The spectrum is flat or negative in the central region of the PWN, while it is positive in the northern and southern regions.
The weighted mean of the spectral index, where the weighting factor is the inverse square of the error, is $-0.041\pm0.014$. The spectral index map shows a kind of band crossing the source from roughly SE to NW with values below $-0.1$ (down to -0.5), where the spectral index error is mostly smaller than 0.3. In the centre of the source the spectral index $\alpha$ is of the order of $-0.2$ with an error ($e_{\rm \alpha,centre}$) of the order of 0.16.
If we take into account the pixels where $e_\alpha$ is at most 0.4, $\alpha$ reaches $0.3\pm0.4$ for the northern part and $0.5\pm0.4$ for the southern part. There is therefore a spectral index variation between the central and northern regions of $|{-0.2-0.3}|=0.5$, which corresponds to 3.1$e_{\rm \alpha,centre}$, and between the central and southern regions of $|{-0.2-0.5}|=0.7$ or 4.4$e_{\rm \alpha,centre}$. However, these are the most extreme variations; in between, the variation is very smooth, as can be seen in Fig.~\ref{FigIndex}.

\begin{figure}
   \centering
   \includegraphics[width=\hsize]{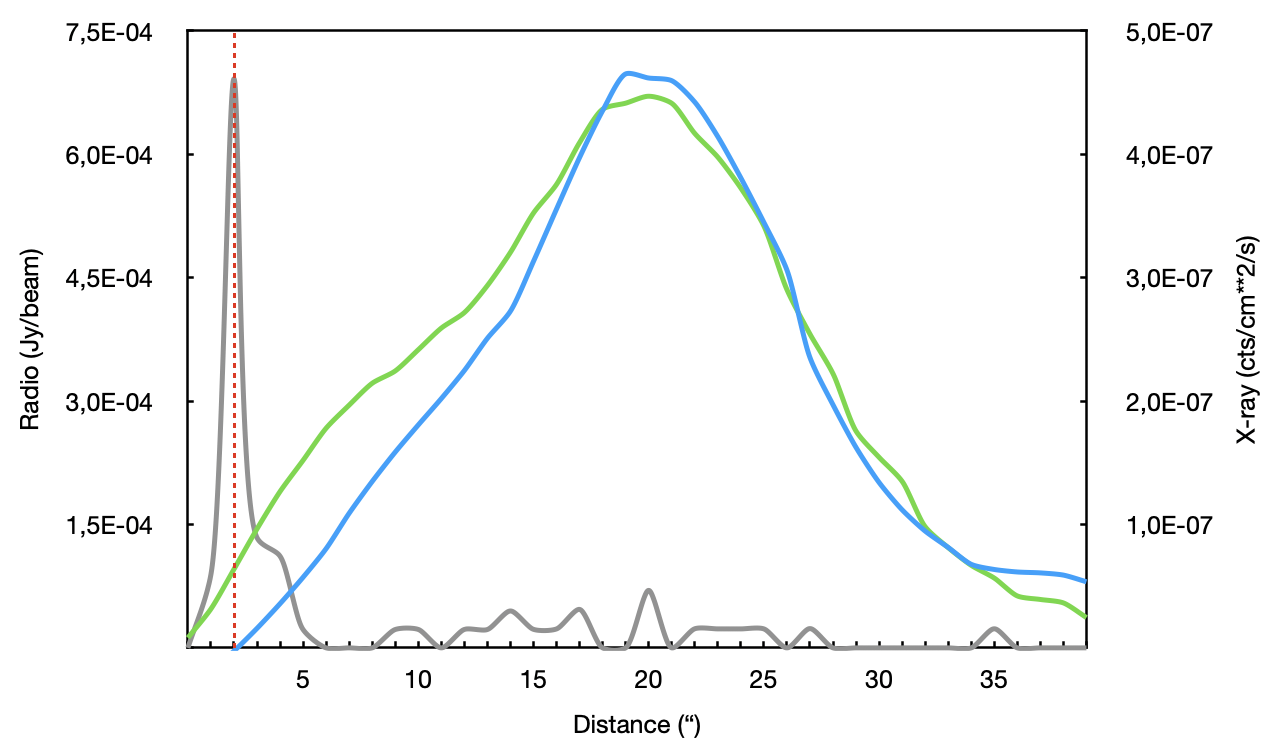}
      \caption{X-ray and radio intensity profiles of PWN PSR\,J2030$+$4415 as a function of the distance from PSR\,J2030$+$4415. The cutting line that defines  the profile scans crosses the pulsar position and follows the proper motion direction of the pulsar (PA$=7.2^{\circ}$). The  grey line  corresponds to 0.5--7 keV X-ray data, whereas the blue line corresponds to the radio data at 736~MHz and the green line to the 1274~MHz data, with the common synthesized beam of $6\arcsec \times 6\arcsec$. The vertical red dashed line marks the pulsar position. We note that the small X-ray peaks far from the pulsar also represent significant emission.}
         \label{FigProfiles}
   \end{figure}

Using the radio images with a common synthesized beam of $6\arcsec \times 6\arcsec$, we   obtained the radio intensity profiles of PWN PSR\,J2030$+$4415 as a function of the distance from the pulsar and compared them with the X-ray profile. The profile scans are shown in Fig.~\ref{FigProfiles}. The X-ray peak emission is coincident with the pulsar position and it is at a distance of $17.7 \arcsec$ from the radio peak. These characteristics are qualitatively similar to those found in PSR\,J1509-5850 (\citealt{Ng2010}, \citealt{Kargaltsev2017}).

In order to estimate the synchrotron radio luminosity, we   first considered from Fig.~\ref{FigIndex} the area of the source where the spectral indices were negative. In this  $20 \arcsec \times 10\arcsec$ area, we   obtained a median value of the negative spectral indices of $-0.23$, with an error $\sim 0.1$, and a  flux density of $1.38\pm0.05$ mJy at 1274~MHz. Integrating the spectrum from 1~GHz to 100~GHz 
and taking a radio spectral index of $-0.23$, we derive a radio luminosity $L_{\rm R, PWN}\approx 1.9 \times 10^{28} d_{0.5 \rm kpc}^{2}$ erg s$^{-1}$. Assuming equipartition arguments
for synchrotron radio sources (\citealt{Pacholczyk1970}),
no protons, an homogeneous leptonic emitter, and that
the magnetic field covers the full volume of the source
taken as a cylinder  $20 \arcsec$ in height and $5 \arcsec$ in radius,
we derive the magnetic field $B_{\rm radio} \sim 93\,\mu$G and a minimum total energy
$E_{\rm tot}  \sim 5\times 10^{41}$ erg for the radio emitter. Future polarization observations using the uGMRT or ngVLA hold the potential to probe the polarization of the extended radio emission trailing the pulsar motion. These observations could provide valuable insights into the magnetic field geometry within the flow.

\subsection{Upper limits for the pulsar and filament}\label{ulim}
There are no hints of the pulsar and the filament at any of the explored radio frequencies. For the pulsar we   obtained a restrictive $3\sigma$ upper limit of 96 $\mu$Jy~beam$^{-1}$ at 736\,MHz  and of 90~$\mu$Jy~beam$^{-1}$ at 1274\,MHz for its radio counterpart. Assuming a typical spectral index of $-1.60$ for the power-law spectra of radio pulsars (\citealt{Jankowski2018}), we converted the $3\sigma$ upper limit of 96 $\mu$Jy~beam$^{-1}$ at 736\,MHz to 1400 MHz, obtaining a value of 34 $\mu$Jy~beam$^{-1}$.
Radio-quiet pulsars are defined as those
that have not been detected in radio down to a flux density limit
of 30 $\mu$Jy at 1400\,MHz (\citealt{Marelli2015}). Therefore, the non-detection of the pulsar in radio provides strong indications of PSR\,J2030$+$4415 belonging to the radio-quiet class of gamma-ray pulsars. For the X-ray filament we can see that the rms noise in the area of $5\arcmin \times 30\arcsec$  covering the first one-third   of the filament from the position of the pulsar is 37~$\mu$Jy~beam$^{-1}$ at 736\,MHz (beam: $4.95'' \times 3.22''$) and 14~$\mu$Jy~beam$^{-1}$ at 1274\,MHz (beam: $2.01'' \times 2.24''$), which means a $3\sigma$ upper limit of 111 ~$\mu$Jy~beam$^{-1}$ and 42~$\mu$Jy~beam$^{-1}$, respectively.

%

%

\section{An apparent near-infrared signature of the PWN}

\begin{figure*}
\begin{center}
\includegraphics[width=\hsize]{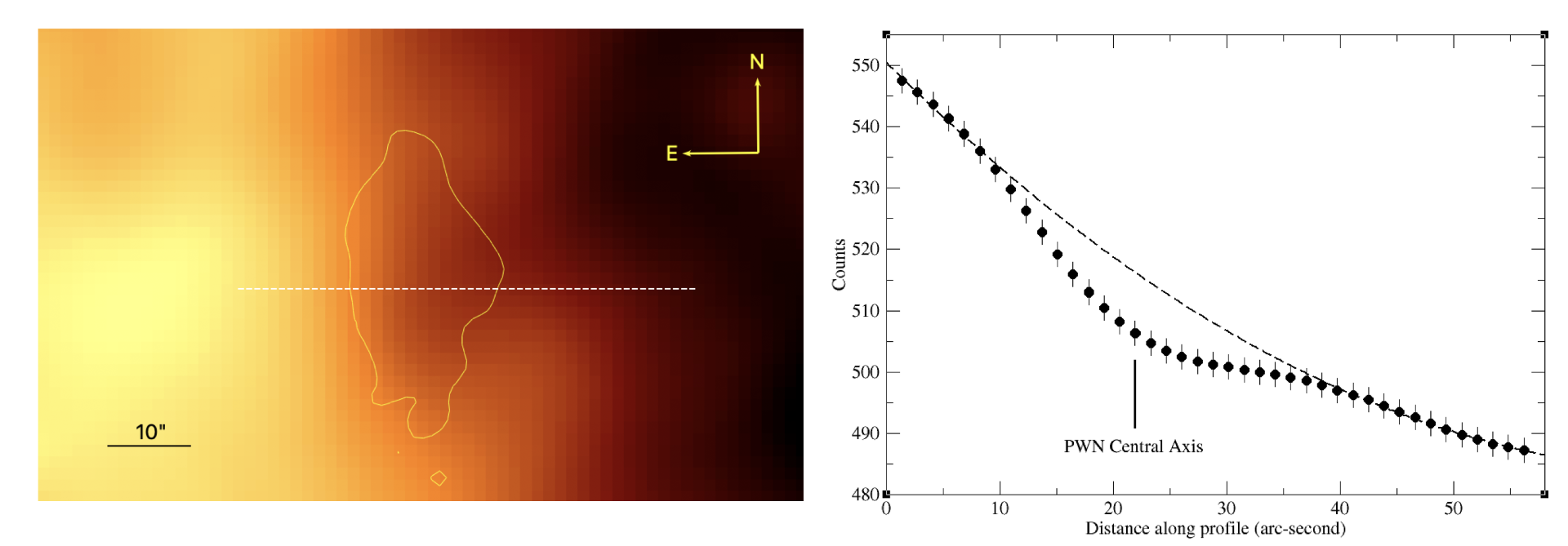}
\end{center}
\caption{Environment of PSR\,J2030$+$4415 using WISE data in the mid-infrared domain. Left: PSR~J2030+4415 region as seen in the 12 $\mu$m W3 band of WISE. The thin yellow line represents the $2\sigma$ contour of the
GMRT 736 MHz radio emission, which matches very closely a minimum of the thermal infrared emission. The yellow compass symbol and the horizontal bar indicate the
image orientation and angular scale, respectively.  Right: East--west emission profile along the white dashed line of the left panel with an attempt to
fit the background emission. There is a  central drop in counts with respect to the dashed line, in coincidence with the PWN central axis.}
\label{WISE}
\end{figure*}

For completeness, we explored the environments of PSR\,J2030$+$4415  using the images
of the Wide-field Infrared Survey Explorer (WISE) in the mid-infrared domain \citep{2010AJ....140.1868W}. It is remarkable that the 12 $\mu$m W3 band view of the region
shows a lack of emission that matches accurately the PWN morphology. This can be seen in the left panel of Fig.~\ref{WISE}, where the
yellow contour outlines the edges of the radio emission from the GMRT 736 MHz map. The contrast has been strengthened in order to highlight the apparent emission drop.
In addition, the right panel of Fig.~\ref{WISE} shows an east--west cut where the emission minimum clearly coincides with the central axis of the PWN.
A simple parabolic profile was fitted, excluding the points overlapping with the GMRT contours, to quantify the significance of the emission drop (about ten counts). Given that the rms in this WISE image amounts to about two counts per pixel, as measured in emission free regions, we are dealing with a likely real 
$\sim 5\sigma$ effect.
The WISE longer wavelength W4 band lacks enough angular resolution
to distinguish this coincidence, while the shorter wavelength bands W1 and W2 are mostly dominated by the stellar content.
It is known that the W3 band  begins to have a significant sensitivity to thermal emission from  interstellar dust.
Therefore, the deficit of its extended emission displayed in Fig. \ref{WISE} could be tentatively interpreted as 
interstellar dust having been swept away by the PWN along its path. If this suggestion is correct, it could open a new aspect of study dealing with the
interplay between pulsar winds and the cold interstellar medium. While examples of optical and infrared emission from bow shocks believed to be associated
with fast moving pulsars exist in the literature
\citep{2006ARA&A..44...17G, 2013ApJ...769..122W}, to our knowledge no previous example of a
dark patch seen against a PWN has been described.

\section{Theoretical interpretation}

The combination of the radio and the X-ray observations of the nebula and the filament associated with PSR\,J2030$+$4415 suggests a physical scenario that can be described as follows: 

(i) Electrons and positrons are mostly accelerated in the region where the pulsar wind terminates, although further shocks can occur downstream of the shocked pulsar wind. 
Most of the acceleration is expected to take place where the dissipation of the unshocked wind energy is the strongest, around the bow-shaped region. There, particles can reach energies of $E=\eta_{\rm acc}qB_{\rm bow}R_{\rm bow}\approx 220\eta_{\rm acc}$~TeV, where $R_{\rm bow}\sim (\dot{E}/4\pi c\rho_{\rm ISM} v_{\rm PSR}^2)^{1/2}\approx 5\times 10^{15}$~cm and $B_{\rm bow}=(\eta_{\rm B}\dot{E}/R_{\rm bow}^2c)^{1/2}\approx 140\,\eta_{\rm B}^{1/2}$~$\mu$G are the size and the magnetic field of the pulsar termination shock region, respectively (taking a medium density of $1$~cm$^{-3}$ and $v_{\rm PSR}=3\times 10^7$~cm~s$^{-1}$, assuming $45^\circ$ between the pulsar motion and the line of sight), and $\eta_{\rm acc}$ is the acceleration efficiency and $\eta_{\rm B}$ the ratio of magnetic to plasma energy density in the shocked pulsar wind, both being $\lesssim 1$.

(ii) Most of the shocked pulsar wind is advected along the tail of the PWN propagating at a significant fraction of $c$ (unless strong mixing with the ISM takes place; see below). Radio synchrotron emission can be produced by the less energetic particles that  follow a hard energy distribution \citep[e.g.][]{Bykov2017}, which explains the hard radio emission of some regions of the PWN. Given the energetics of the forward shock, $\lesssim v_{\rm PSR}\,\dot{E}/c\approx 3\times 10^{31}$~erg~s$^{-1}$, and its high temperature, $kT\sim (3/32)v_{\rm PSR}^2\sim 100$~eV for $v_{\rm PSR}=3\times 10^7$~cm~s$^{-1}$, thermal radio emission from the shocked ISM seems insufficient to explain the detected radio fluxes. 

(iii) The softer radio emission found at $\sim20\arcsec$ from the pulsar may be attributed to a Mach disk (re-collimation) shock produced by the medium thermal pressure, which should form roughly at that distance for typical ISM values: $r_{\rm Mach}\approx (\dot{E}/4\pi cP_{\rm ISM})^{1/2}\approx 2\arcsec$ at 0.5~kpc \citep[where the thermal pressure of the ISM, $P_{\rm ISM}\sim 1$~eV~cm$^{-3}$, was adopted; e.g.][]{Barkov2019b}. In this complex region, particle acceleration may differ from that at the pulsar wind termination, and non-thermal particle energy may be redistributed towards lower energies, which  explains the steeper radio spectrum. Entrainment of shocked ISM may also favour this redistribution by reducing the averaged particle energy. 

(iv) The X-ray emission from the tail, of likely synchrotron origin, comes from a softer region of the particle spectrum than the radio emission \citep[e.g.][]{Bykov2017}. The observed steepening with distance from the pulsar, with a softening of the X-ray emission along the PWN of $\Delta  \Gamma_{X} = 0.8 \pm 0.3$  has been interpreted \citep{Vries2020} as a consequence of synchrotron cooling of the electrons and positrons emitting those photons,
 for example with energies of $\sim 30$~TeV for a shocked PWN magnetic field $B_{\rm PWN} \sim 30$~$\mu$G. This cooling takes place on a timescale of a few hundred  years, whereas the shocked wind should be leaving the region on   a    timescale of a few years unless strongly slowed down from mildly relativistic velocities down to hundreds of km s$^{-1}$, which may occur if strong mixing with the ISM takes place \citep[e.g. due to pinching or Kelvin-Helmholtz instabilities;][]{Barkov2019,Vries2020}. However, energy-dependent diffusive escape into the ISM by the emitting particles is much faster than synchrotron cooling, and thus may be the actual mechanism behind the observed X-ray steepening as these particles should diffuse out of the region on a timescale of a  few~years. Taking into account that the slowest diffusion corresponds to the Bohm regime, the diffusion length of particles producing $\sim 1$~keV synchrotron photons is $\gtrsim 5\arcsec\,E_{\rm 30TeV}^{1/2}\,t_{\rm yr}^{1/2}\,B_{30\mu{\rm G}}^{-1/2}$, comparable to the width of the structure, so on those scales the particle speed would be close to $c$. The lack of detection of X-rays from the escaped particles outside the nebula tail could be a consequence of a significantly weaker ISM magnetic field, which implies a much lower synchrotron emissivity and larger diffusion length, both yielding a low surface brightness.

(v) The obtained mean radio spectral index of the PWN is $\alpha_{\rm R, median}=-0.041\pm0.014$. The X-ray emission is also described by a power law, with a photon index $\Gamma \equiv 1-\alpha_{\rm X}=1.48\pm0.10$. The extrapolation of the radio power law to X-ray frequencies clearly shows that the X-ray emission is several orders of magnitude lower. Because of this, it can be inferred that the energy particle distribution should follow a broken power law, harder at lower energies. This feature of the energy distribution is not related to the diffusive escape of the particles, which only affects the highest energy end of the distribution.

vi) Predictions in gamma rays can be done assuming that the highest energy electrons and positrons escaped from the nebula interact via inverse-Compton (IC) scattering with ambient IR/optical photons. Adopting a typical IC target energy densities of 1~eV~cm$^{-3}$, an emitting region size of $\sim 10'$, Bohm diffusion, and a power in particles of $\sim 0.1\,\dot{E}$, the IC flux may reach $\sim 10^{-13}$~erg~s$^{-1}$~cm$^{-2}$. We note however that this is an optimistic prediction as particles may diffuse out of the region much faster. 

vii) As the pulsar moves, ordered ISM magnetic lines are entrained and dragged by the shocked ISM, subsequently becoming amplified with respect to the surrounding ISM magnetic field, due to the interaction. The most energetic particles accelerated in the pulsar wind termination region can reach these ordered magnetic lines and escape into the ISM, where they form filamentary structures that shine in X-rays via synchrotron emission \citep[e.g.][]{Barkov2019,Olmi2024}. Adopting a reference value of $\sim$10~$\mu$G for the filament magnetic field, 1~keV photons are produced by $\approx 40\,B^{-1/2}_{10\mu{\rm G}}$~TeV particles, with an associated radiative time of $t_{\rm sync}\sim 3\,B^{-3/2}_{10\mu{\rm G}}$~kyr. Given that the magnetic field seems to be mostly ordered and the X-ray filament is narrow, particles are likely to cross the 15$\arcmin$ (2.2 pc) filament at almost the speed of light, which yields a crossing of time $t_{\rm cross}\sim 7$~yr. Given the observed filament X-ray luminosity of $L_{\rm Fil,X}\approx 3.6\times 10^{30}$~erg~s$^{-1}$ and the involved timescales, the minimum injected particle luminosity should be $\sim (t_{\rm sync}/t_{\rm cross})L_{\rm Fil,X}\sim 1.5\times 10^{33}$~erg~s$^{-1}$, which is   $\sim$10\%  of the $\dot{E}$. We note that the $\sim 4\arcsec$ width of the filament is compatible with the $\sim 10\arcsec$ scale of the particle gyroradii.

The large-scale X-ray filament, thought to be of synchrotron origin, may have a radio counterpart that has not yet been detected. Our GMRT observations allow us to derive a lower limit on the energy of the particles escaping the bow-shaped region around the pulsar, for which we followed \cite{Bordas2021}. First, we computed the expected synchrotron radio luminosity for particles being injected at the base of the filament. For that, we assumed a total jet injection power of $L_{\rm inj} = 1.5 \times 10^{33}$~erg~s$^{-1}$. Particles are injected following a power-law distribution with exponential index of 2.2, featuring a cut-off at  its low- and high-energy ends. For the low-energy cut-off, we considered values ranging from $\gamma_{\rm min}^{\rm cut} = 10^{2}$ to $10^{5}$, whereas we fixed the high-energy to $\gamma_{\rm max}^{\rm cut} = 10^{8}$. We scanned different values of the filament magnetic field $B_{\rm Fil} = 10-100$~$\mu$G, which are higher than those typically found in the ISM as they may be generated by an amplification produced by the relativistic particles themselves (see e.g. \citealp{Olmi2024}). The radio synchrotron fluxes of the filament were then computed and compared to the upper limits obtained with GMRT Band 4 and Band 5, accounting for the instrument beam at each band, with the additional constraint to provide a luminosity in the X-ray band at the level of the observed luminosity, $L_{X} \sim 4 \times 10^{30}$~erg~s$^{-1}$ \citep{Vries2020}. The most conservative lower limit for $\gamma_{\rm min}^{\rm cut}$ found in the $B_{\rm Fil}$ value range explored is of a few times $10^3$;  below that value our GMRT observations should have detected the radio emission.

\section{Summary and conclusions}

We have reported on the discovery of the likely radio counterpart of the PWN associated with PSR J2030$+$4415 using GMRT dedicated observations of the source at 736 MHz and 1274 MHz frequencies. The extended radio emission is trailing the pulsar motion, with the radio peak found at 0.045 pc from the pulsar location (assuming a distance to the source of 0.5 pc). In X-rays the emission peaks at the pulsar position, and its extension is shorter than in radio. Both radio and X-ray emission might be produced by the same particle population, although the former requires a harder injection index, implying a softening break in the parent particle spectrum. This softening could be interpreted as radiative cooling in the PWN due to synchrotron losses. However, this  would require that the flow and the ultrarelativistic particles behind the X-rays remain within the PWN region for much longer times than the flow advection time. Alternatively, the observed X-ray softening would be due to high-energy particles escaping the PWN region through energy-dependent diffusion. Such a scenario has been used to successfully explain a number of large-scale features from bow-shock PWNe (see e.g. \citealt{Bandiera2008}). In the case of PSR~J2030$+$4415, the source indeed displays   an X-ray jet-like structure extending unbent for a few parsecs into the surrounding ISM. Our GMRT observations did not reveal a radio counterpart of this jet-like structure, which places a constraint on the cut-off energy for particles being able to escape the PWN at a $\gamma_{\rm min}^{\rm cut}$ of a few $\times 10^3$. Additionally, we report observations of the source using images of WISE in the mid-infrared band. The emission displays a cavity which spatially coincides with the PWN location. This suggests a novel method to study the interaction of the PWN and the dust present in the surrounding medium.

\begin{acknowledgements}
       
We thank the anonymous referee for useful suggestions and comments that helped to improve the content of the manuscript. We thank the staff of the GMRT that made these observations possible. GMRT is run by the National Centre for Radio Astrophysics of the Tata Institute of Fundamental Research. The authors would like to thank Dr. D. B. Palakkatharappil for providing a code for automatic alignment of GMRT images. We (JMP, VB-R, PB) acknowledge financial support from the State Agency for Research of the Spanish Ministry of Science and Innovation under grants PID2022-136828NB-C41/AEI/10.13039/501100011033/ERDF/EU, PID2022-138172NBC43/AEI/10.13039/501100011033/ERDF/EU and through the Unit of Excellence María de Maeztu 2020-2023 award to the Institute of Cosmos Sciences (CEX2019-000918-M). We acknowledge financial support from Departament de Recerca i Universitats of Generalitat de Catalunya through grant 2021SGR00679. JM acknowledges support from Junta de Andaluc\'{\i}a under Plan Complementario de I$+$D$+$I (Ref. AST22\_00001), Plan Andaluz de Investigaci\'on, Desarrollo e Innovaci\'on as research
group FQM-322, Spanish Ministerio de Ciencia e Innovaci\'on,  
Agencia Estatal de Investigac\'on (Ref. PID2022-136828NB-C42)
and FEDER funds. Part of the scientific results reported in this article are based on observations made by the Chandra X-ray Observatory and published previously in cited articles [Chandra ObsId 14827, 20298, 22171-22173, 23536, 24954, 24236].
This research has made use of NASA’s Astrophysics Data System. This research has made use of the SIMBAD database, operated at CDS, Strasbourg, France. This publication makes use of data products from the Wide-field Infrared Survey Explorer, which is a joint project of the University of California, Los Angeles, and the Jet Propulsion Laboratory/California Institute of Technology, funded by the National Aeronautics and Space Administration. V.B-R. is Correspondent Researcher of CONICET, Argentina, at the IAR.

\end{acknowledgements}

\bibliographystyle{aa}
\bibliography{references.bib}{}

\end{document}